\documentclass[12pt]{spieman}  
\usepackage[]{graphicx}
\usepackage{setspace}
\usepackage{tocloft}
\usepackage{epsfig}
\usepackage{amsmath}
\usepackage{natbib}
\usepackage{color}
\usepackage{lscape}

\usepackage{deluxetable}
\usepackage{longtable}
\usepackage{rotating}

\def\apj{ApJ}

\def\mnras{MNRAS}
\def\nat{Nature}

\def\aap{A$\&$A}
\def\apjl{ApJ}

\def\pasp{PASP}

\title{First exoplanet transit observation with the Stratospheric Observatory for Infrared Astronomy: Confirmation of Rayleigh scattering in HD 189733 b with HIPO} 

\author{Daniel Angerhausen,\supscr{a,b} Georgi Mandushev,\supscr{c} Avi Mandell,\supscr{a} Edward Dunham\supscr{c}, Eric Becklin,\supscr{d,e},Peter Collins,\supscr{c} Ryan Hamilton,\supscr{e} Sarah E. Logsdon,\supscr{d} Michael McElwain,\supscr{a} Ian McLean,\supscr{d} Enrico Pf\"uller,\supscr{f} Maureen Savage,\supscr{e} Sachindev Shenoy,\supscr{e} William Vacca,\supscr{e} Jeff Van Cleve,\supscr{e} J\"urgen Wolf\supscr{f} }

\affiliation{\supscrsm{a}Exoplanets and Stellar Astrophysics Laboratory, Code 667, NASA Goddard Space Flight Center, Greenbelt, MD 20771, USA\\
\supscrsm{b}Rensselaer Polytechnic Institute (RPI), 110 Eighth Street, Troy, NY USA 12180\\
\supscrsm{c}Lowell Observatory, 1400 West Mars Hill Road, Flagstaff, AZ 86001, USA\\
\supscrsm{d}Department of Physics and Astronomy, University of California Los Angeles (UCLA), 465 Portola Plaza, Los Angeles, CA 90095, USA\\
\supscrsm{e}USRA-SOFIA Science Center, NASA Ames Research Center, Moffett Field, CA 94035, USA\\
\supscrsm{f}Deutsches SOFIA Institut, University of Stuttgart Pfaffenwaldring 29, D-70569 Stuttgart, Germany
}

\cftpagenumbersoff{figure}
\cftpagenumbersoff{table} 
\begin{document} 
\maketitle 

\begin{abstract}
Here we report on the first successful exoplanet transit observation with the Stratospheric Observatory for Infrared Astronomy (SOFIA). We observed a single transit of the hot Jupiter HD~189733 b, obtaining two simultaneous primary transit lightcurves in the B and $z^\prime$ bands as a demonstration of SOFIA's capability to perform absolute transit photometry. We present a detailed description of our data reduction, in particular the correlation of photometric systematics with various in-flight parameters unique to the airborne observing environment. The derived transit depths at B and $z^\prime$ wavelengths confirm a previously reported slope in the optical transmission spectrum of HD~189733 b. Our results give new insights to the current discussion about the source of this  Rayleigh scattering in the upper atmosphere and the question of fixed limb darkening coefficients in fitting routines.
\end{abstract}


{\noindent \footnotesize{\bf Address all correspondence to}: Daniel Angerhausen, Exoplanets and Stellar Astrophysics Laboratory, Code 667, NASA Goddard Space Flight Center, Greenbelt, MD 20771, USA; Tel: +1 301.286.0454; E-mail:  \linkable{daniel.angerhausen@nasa.gov} }

\begin{spacing}{2}   

\section{Introduction}
The exploration and detailed analysis of exoplanet atmospheres is one of the most dynamic fields of astrophysics today. The first  successful observations probing the atmospheric properties of Jupiters were conducted with the Hubble Space Telescope's (\textit{HST}) STIS instrument \citep{2002ApJ...568..377C} and with the Spitzer Space Telescope's (\textit{Spitzer}) IRS instrument \citep{2007Natur.445..892R, 2007ApJ...658L.115G}. Approximately a decade later, we are now able to analyze the atmospheres of planets down to Neptune \citep[GJ~436b,][]{2014Natur.505...66K} and super-Earth \citep[GJ~1214b,][]{2014Natur.505...69K} sizes.   High signal-to-noise observations of the absorption depth of molecular bands with strong signatures in the optical and infrared portions of the spectrum can determine both the temperature-pressure thermal profile as well as the abundances of atmospheric constituents, helping to constrain the overall atmospheric chemistry and structure \citep{2008ApJ...678.1436B,2008ApJ...683.1104F,2012ApJ...758...36M}. It has also been shown that the relative abundances of atomic species such as C, O, and N in a gas giant's atmosphere could be indicative of the region of the proto-planetary disk in which the planet formed \citep{2004ApJ...611..587L,2011ApJ...727...77M,2011ApJ...743..191M}.

In this paper we present the first observations of a transiting planet, the well-known transiting Hot Jupiter HD~189733 b, using the Stratospheric Observatory for Infrared Astronomy  \citep[SOFIA,][]{2007SPIE.6678E...8B,2012ApJ...749L..17Y}.  SOFIA consists of a 2.5-meter telescope mounted within a modified Boeing 747-SP aircraft operating at altitudes up to 45,000 feet, thereby offering the opportunity for observations at altitudes where the telluric absorption from the Earth's atmosphere is greatly reduced.  In Section \ref{chap:obs} we describe our observations, in Section \ref{chap:red} we describe our data reduction and transit light curve fitting analysis, in Section \ref{chap:res} we describe our results, and discuss the implications for the atmosphere of HD~189733~b and for future observations with SOFIA in Section \ref{chap:sum}.

\subsection{Exoplanet observations with SOFIA} 

As an airborne observatory, SOFIA has a number of potential advantages for precise time-domain spectrophotometric observations of transiting exoplanets.  Ground-based observations are significantly affected by variations of absorption from telluric gases, in particular $H_2O$, in the Earth's atmosphere, and these same gases are also the species of interest  in exoplanet atmospheres; SOFIA can observe in important atmospheric windows not observable from the ground \citep{2003ASPC..294..591S,2007ASPC..366..256D, 2010PASP..122.1020A,2010PhDT.......210A}. These are mostly the water bands but also $CO$, $CH_4$, $CO_2$ are much better mixed and therefore reduce the temporal variation in these bands, which is a crucial point for time-series observations. These bands are also the molecular bands of interest in targets such as HD 189733 b.  The airborne observatory operates in the wavelength regime where the planet's black-body temperature peaks and contrast ratios between star and planet improve, and the SOFIA telescope also operates at lower temperatures (240K) than ground-based telescopes; therefore the contribution from the thermal background (the dominant noise source for transit observations at wavelengths longer than 3 microns) are significantly reduced.  SOFIA can observe simultaneously at infrared and  optical wavelengths using its FLITECAM \citep{2006SPIE.6269E.168M} and HIPO \citep{2004SPIE.5492..592D} instruments in 'FLIPO' mode \citep{2012SPIE.8446E..18D}, thereby obtaining light curves for a single transit event over a wide range in wavelength.

However, there are also certain challenges when observing with SOFIA due to the airborne observing environment. In test flights HIPO photometry has been shown to be affected by terrestrial Rayleigh scattering, by ozone extinction in the Chappuis band that varies with position along the flight path, potentially by volcanic aerosols, by pointing errors, and by other factors affecting the PSF \citep{ 2014SPIE.9147E..0HD}.  At optical wavelengths the PSF is dominated by wavefront aberrations imposed by the turbulent shear layer that passes over the telescope cavity.  The strength of the shear layer density fluctuations varies with the static air density, causing density-dependent interaction of the very broad wings of the PSF with the photometric aperture.  Furthermore, the logistics of aircraft operations result in observation windows which are limited in duration and dependent on the flight plan chosen for a specific flight; this can result in the limited availability of measurements before or after a transit event and hamper the decorrelation of these various observational effects with the transit light curve.

Fortunately an expected dependence on Mach number has not been seen, and the residual systematic noise due to impacts of focus errors and high speed image jitter are much less than the photon noise if large circular synthetic photometric apertures are used (see also discussion in \ref{fig:comp_noise}).  Other potential sources of systematic error that have not yet been investigated include higher-order extinction corrections, water vapor absorption in the $z^\prime$ band, possible polarization effects from the SOFIA tertiary mirror, and the impact of a known temperature-dependent astigmatism term in the PSF.  We have also not yet investigated whether weighted aperture photometry is beneficial for precise photometric work. Furthermore, Rayleigh extinction is well behaved in flight because the aircraft flies at constant pressure (and therefore constant Rayleigh zenith optical depth) and the usual extinction correction can be applied.  Ozone extinction can largely be avoided by proper filter selection and volcanic aerosols are thankfully uncommon.  Pointing and focus errors, and variations with instrument and environmental factors must be decorrelated using SOFIA housekeeping data, to reach the high signal to noise needed for our differential transit depth measurements.

\subsubsection*{HD 189733 b}

HD 189733 b \citep{2005A&A...444L..15B} is a Jupiter-mass planet orbiting at 0.03 AU around one of the 
closest K-type stars; the deep transit signal in addition to a very bright, nearby host star 
results in the best opportunity for high-precision characterization of any known exoplanet. This 
transiting system is a benchmark for exoplanet observations and has been the target for 
many ground- and space-based observations. Both multi-band photometry and spectroscopy 
with \textit{Spitzer} have provided measurements of the mid-IR emission from the planet by measuring 
the occultation of the planet by the central star, probing both molecular absorption and the 
temperature structure in the bulk of the upper atmosphere \citep{2008ApJ...686.1341C, 2008Natur.456..767G}. Observations with \textit{HST} have explored molecular bands in the NIR, and indeed, early 
results with the NICMOS spectrograph claimed absorption from $H_2O$ \citep{2007Natur.448..163V,2014ApJ...791...55M}, $CH_4$ \citep{2008Natur.452..329S} and $CO_2$ between 1.5 
and 2.5 $\mu$m . Other observations found the same molecular features in the emission spectra observed during secondary eclipse observations \citep{2007ApJ...658L.115G,2009ApJ...690L.114S}. However, there has been considerable discussion on this topic and the quality and reproduction of these results \citep[e.g.,][]{2011MNRAS.411.2199G, 2011ASPC..450...63D, 2014ApJ...784..133S}. It is still uncertain whether these molecular features exist or if there is a haze that obscures wavelengths below 2 $\mu$m \citep{2008MNRAS.385..109P,2009A&A...505..891S}, in fact, \cite{2013MNRAS.432.2917P} make the argument that HD 189733 b's atmosphere is most likely dominated by Rayleigh scattering in the visible and near-infrared. While \citep{2007A&A...476.1347P} and other observations did not directly detect any starspots by crossing events, a recent study by \citep{2014ApJ...791...55M} argued that the measured slope at shorter wavelength could also partly be caused by unocculted star spots, reducing the contribution of Rayleigh scattering to molecular hydrogen and not necessarily dust in HD 189733 b's atmosphere.

Our SOFIA observations were designed to both examine the presence of a strong Rayleigh slope in the optical (using photometric observations in B and $z^\prime$ filters with HIPO) and either confirm or reject the existence of absorption from $H_2O$ in the NIR (using photometric measurements in FLITECAM's Paschen alpha  1.88 $\mu$m filter, just longwards of the current upper limit of  HST and unobservable from the ground). While both of these features were already measured in various observations with HST WFC3 and NICMOS (see references above) they still added a compelling scientific value (reproduction and direct comparison) to our observations, which were mainly a proof of concept experiment for SOFIA. These observations leverage the advantages of SOFIA for simultaneous optical and near-infrared observations that are difficult or impossible from the ground, as well as providing an optimal target for initial tests of the precision possible for exoplanet transits with SOFIA.

\section{Observation}\label{chap:obs}

We observed HD 189733 b during a transit on SOFIA's flight  number 134 on UT Oct 1 2013 as part of a Cycle 1 GO program (PI: Mandell, Proposal ID: 01-0099).  Observations were conducted in the FLIPO configuration (FLITECAM and HIPO operating simultaneously) in order to observe in three optical and infrared bands at the same time: B and $z^\prime$ with HIPO and a narrow-band filter covering the Paschen $\alpha$ spectral feature at 1.88 $\mu$m  with  FLITECAM. The HIPO filters were  selected to avoid spectral regions with potentially high ozone variability, while the FLITECAM filter was chosen due to its wavelength coverage of a prominent $H_2O$ spectral feature that cannot be sampled from ground-based observatories. An additional optical channel (for general calibration or tracing of a specific telluric absorption band) can be obtained from the Focal Plane Imager (FPI+), though we were unable to acquire FPI+ data for our flight due to an instrument malfunction (see below).

Due to constraints on the flight plan imposed by requirements on the direction and timing of SOFIA flights, the observing period only allowed for a very short baseline before ingress and almost no baseline after egress. Two different instrument operators were running the blue and red sides and due to some problems with the general acquisition of stable condition for observing (see below), the blue side started later than the red side. In detail we got  0.00 hours before, 0.20 hours after, and 1.77 hours of in-transit for the blue side and 0.26 hours before, 0.20 hours after, and 1.76 hours of in-transit for the red side. The duty cycle for both HIPO channels is essentially 100 percent with a minimal dead time due to the CCD frame transfer of a few milliseconds, because in the frame transfer CCD the readout is overlapped with the next integration.

Various instrumental issues emerged before and during the flight that further hampered the acquisition of high-quality data. In mid September 2013, several components in the FLITECAM detector computer failed and were replaced by spares. At the same time, a required change to the liquid helium venting configuration led to the creation of a thermo-acoustic oscillation that reduced the liquid He hold time by approximately 50\%. Flights were modified to account for the shorter Helium hold time, but unfortunately, the replacement computer components led to improper initialization of the FLITECAM detector. In the case of observations in which we simply stare at the source, or observation sequences utilizing coadds, this quickly resulted in source saturation. The problem was not noticed until the conclusion of the September/October 2013 flight series and the start of data analysis in mid October 2013. The Helium hold time as well as the electronics problem have been fixed in the meantime, and we successfully demonstrated FLITECAM observations of bright targets during FLIPO commissioning and early science in Feb 2014 \citep{2014SPIE.9147E..2YL}.

In order to capture the light that is passed through the SOFIA telescope's dichroic tertiary mirror \citep[25\% and 45\% reflectivity for the B and z' bandpasses; see][]{2012SPIE.8446E..18D} to the Focal Plane Imager (FPI+) we had planned to save its image data for scientific analysis. Most of the visual light passes the tertiary beam splitter before it is reflected into the Nasmyth tube by the fully-reflective tertiary.
The peak transmission of the tertiary beam splitter is at 570 nm with 64\% of the light passing the mirror. A significant amount of visual light is not transmitted but rather absorbed or reflected along with the longer, infrared wavelengths. However, in the range between 480 nm to 800 nm where the visual-light CCD cameras are most sensitive, more than 50\% of the light is transmitted. The FPI+ contains a highly sensitive and fast EM CCD camera. Its images are primarily used for tracking but can also be stored without disrupting the tracking process and in parallel with measurements of the instruments mounted to the telescope. Unfortunately, the FPI+ suffered a failure of its controller electronics right at the beginning of the exoplanet observation leg, forcing us to abandon FPI+ data taking and tracking. 

A faulty power supply in the controller electronics was found to be the cause of the failure. Since its replacement the FPI+ has worked reliably during future flights, including a mission for an exoplanet transit observation in early 2014 (PI: Angerhausen $\&$ Dreyer, in prep.), where we were able to characterize its photometric performance. Starting with SOFIA observing cycle 4 (2015), the FPI+ will be made available for proposals as a facility science instrument.

The HIPO instrument was operated in Basic Occultation mode with full-frame read-out to maximize our field standard possibilities, though
we later concluded that the available field standards were too faint to use for differential photometry measurements. Our final HIPO data set consists of about $\sim$2700 measurements of 3 s integrations on red side in the Sloan $z^\prime$ filter and $\sim$1000 measurements of 7 s integrations on the blue side in the Johnson B filter. As a back-up for the FPI, SOFIA's Fine Field Imager (FFI) was used for tracking. Due to its lower spatial resolution and defocussing during the cool-down phase, tracking accuracy was reduced.  The time required to recover from the FPI+ failure and switch to the FFI also caused nearly all of the pre-transit baseline to be lost.  This loss was particularly acute because the flight plan, scheduled late in the observing season, included no post-transit baseline, leaving us in the difficult position of analyzing a transit with no out-of-transit baseline for noise characterization.


\section{Data reduction}\label{chap:red}

\subsection{Image processing}
Due to the failure of FLITECAM and FPI only the HIPO data was reduced. The raw HIPO CCD frames were bias-subtracted and overscan-corrected.  Flat fielding was complicated by the fact that the shape of twilight sky flats did not match the behavior of moving star images.  We suspect that this was due to a stray light issue, possibly due to insufficient baffling or due to the reflective conical ``button" installed in the central obscuration of the SOFIA telescope to reduce FLITECAM's thermal background.  In the B filter the 1-2\% amplitude laser annealing pattern common in e2v CCDs was apparent, and in both filters a pattern similar to vignetting was seen in the sky flats that was not seen in the positional dependence of star flux.   Our final flat fielding approach for the B filter data was to fit the overall curvature with a low-order polynomial and divide it in the twilight flats leaving the laser annealing pattern intact but removing the overall curvature.Flat field correction was done with this modified flat. The removal of the low order curvature was necessary to isolate the low-level annealing patterns. Otherwise the flatfielding will be dominated by the high-flux sky flat and introduce a bias. The flatfielding then removes the annealing pattern.  On the red side we bypassed flat field correction altogether.  We are currently working on lab tests to understand and improve the flat fielding procedure for future HIPO observations.

%

\subsection{Aperture photometry}

We performed aperture photometry using large aperture diameters (60 arcsec) to reduce the effects of focus, jitter, and tracking errors. For our very bright target, sky and read noise contributions were each only $\sim10\%$ of the stellar photon noise even in this very large aperture (see also section \ref{fig:comp_noise}). We avoided PSF fitting because the shear layer PSF is not well represented analytically. However we did use PSF fitting to estimate the brightness ratio of the target and its faint late-type companion and to derive the characteristics of the PSF shape (i.e. FWHM, ellipticity) for each exposure. The total number of electrons per exposure is $5.81 \times 10^6$ in the B channel, and  $1.44\times 10^7$ in  $z^\prime$. The total number of sky electrons in this aperture is 
$1.46\times 10^5$ for B, and $2.83\times 10^6$ in $z^\prime$. See section \ref{fig:comp_noise} for a detailed noise budget.

The initially derived transit depths were diluted due to the inclusion of the fainter companion star (HD 189733 B, M4V, d$\sim12"$, V$\sim10 mag$) in our photometric aperture. The dilution was undetectable in the B filter and $\sim1.5\%$ in the $z^\prime$ filter.  Propagating this error into the final radius ratio, the correction for $R_p/R_*$ should be 0.0012 at $z^\prime$ (see Table \ref{tbl:red_res}, footnotes). any other stars in the aperture are
unmeasurable in z'. In B neither companion is measurable.

\subsection{Decorrelation and transit fitting}\label{noisemod}

\begin{figure*}[]
  \centering
      \includegraphics*[width=0.9\textwidth]{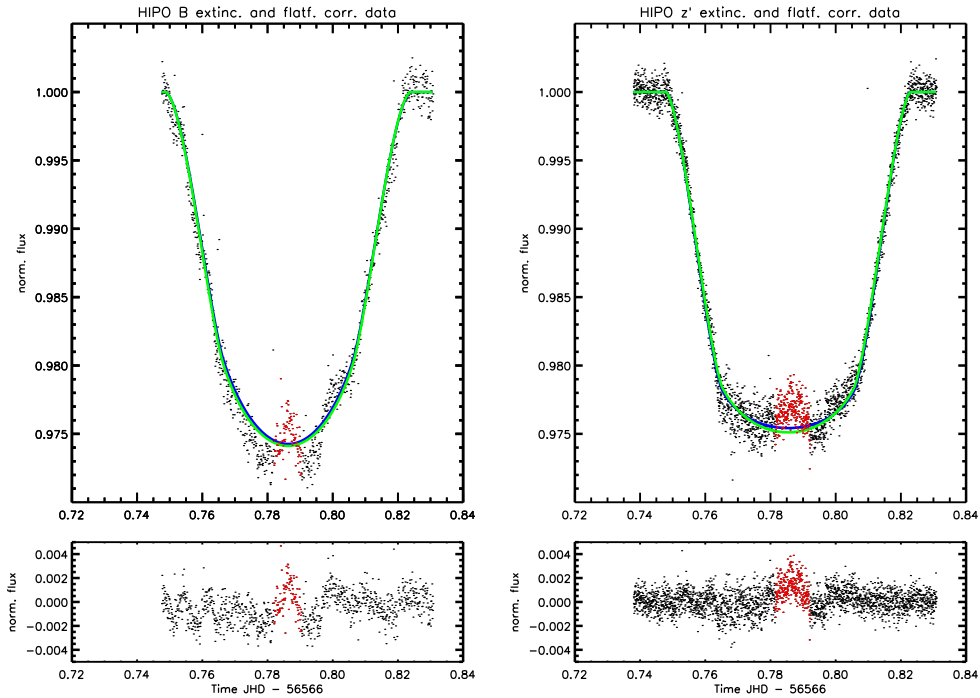}
       \caption{Extinction and flat field corrected absolute photometry of HIPO's B (left) and $z^\prime$ (right) channels before removal of correlated noise: the transit is clearly visible with high confidence -- instant evidence that SOFIA is indeed a great platform for photometric observations of this kind. For this first iteration fit (blue) the ``starspot" region (red) was excluded. Green: an alternative fit with fixed limb darkening.}
     \label{fig:1ord}
\end{figure*}

The absolute raw lightcurves obtained with aperture photometry clearly show a transit signature (see Figure \ref{fig:1ord}). Correction of flux variations using the nearby field stars was not necessary.  However, the time series of the comparison stars were added to the base parameter set for the following analysis in order to allow for this correction if significant on a lower level. Table \ref{tab:corr1} lists the correlation of all parameters and shows that the effect of the comparison stars is negligible. Other observational parameters such as the focus, overscan or air density have a much larger correlation: An initial inspection of the light curves revealed a significant amount of this correlated ('red') noise in the raw photometry, such as a short increase in flux during the transit, with a shape similar to the deviation that would be caused by the planet transiting a cooler starspot \citep{2014MNRAS.442.3686B,2014IAUS..293..435A}. To avoid bias from this feature in our initial light curve fitting, we fit the light curves with the signature masked out. Absent out-of-transit baseline data we attempted to correct for the correlated noise by combining an analytic model for the exoplanet transit light curve \citep{2002ApJ...580L.171M} with a linear combination of normalized (through division by their mean) observational parameters $p_i(t)$, sampled at the same time as our exposures, as a model of the residuals ($R_{model}(t)=\sum c_i \times p_i(t)$). For the light curve model, we fixed the planetary orbital parameters to values from previous observations \citep[see Tables \ref{tbl:blue_res} and \ref{tbl:red_res}]{2008ApJ...677.1324T} but left the transit depth and stellar limb darkening (optional) as free parameters. The model parameters were optimized through a multiple linear regression analysis, and this noise model was then subtracted from the original data. Table \ref{tbl:corr1}  shows the (linear Pearson) correlation coefficients of each observational parameter with the respective residuals as well as the mean values that were used for normalization.

This noise model was computed independently from the later MCMC analysis (see section \ref{section:MCMC}) and no marginalization has been done over the instrument model correlation slope terms. We choose this approach because the lack of baseline would have caused convergence issues and a high risk of running into degeneracies between the noise model and the actual transit depth. 
For the analysis of the cycle 2 data with more baseline and house keeping data we are working on various integrated approaches using either Gaussian processes or a marginalization over the parameter regression. For this analysis we estimate the maximum errors introduced very conservatively by calculating the difference between the standard deviations before and after correction (see Table \ref{tbl:noisecomp}) and using it to derive the outer error bars in Figure \ref{fig:comp_res}.

 \begin{figure*}[htp]
  \centering
      \includegraphics*[width=0.98\textwidth]{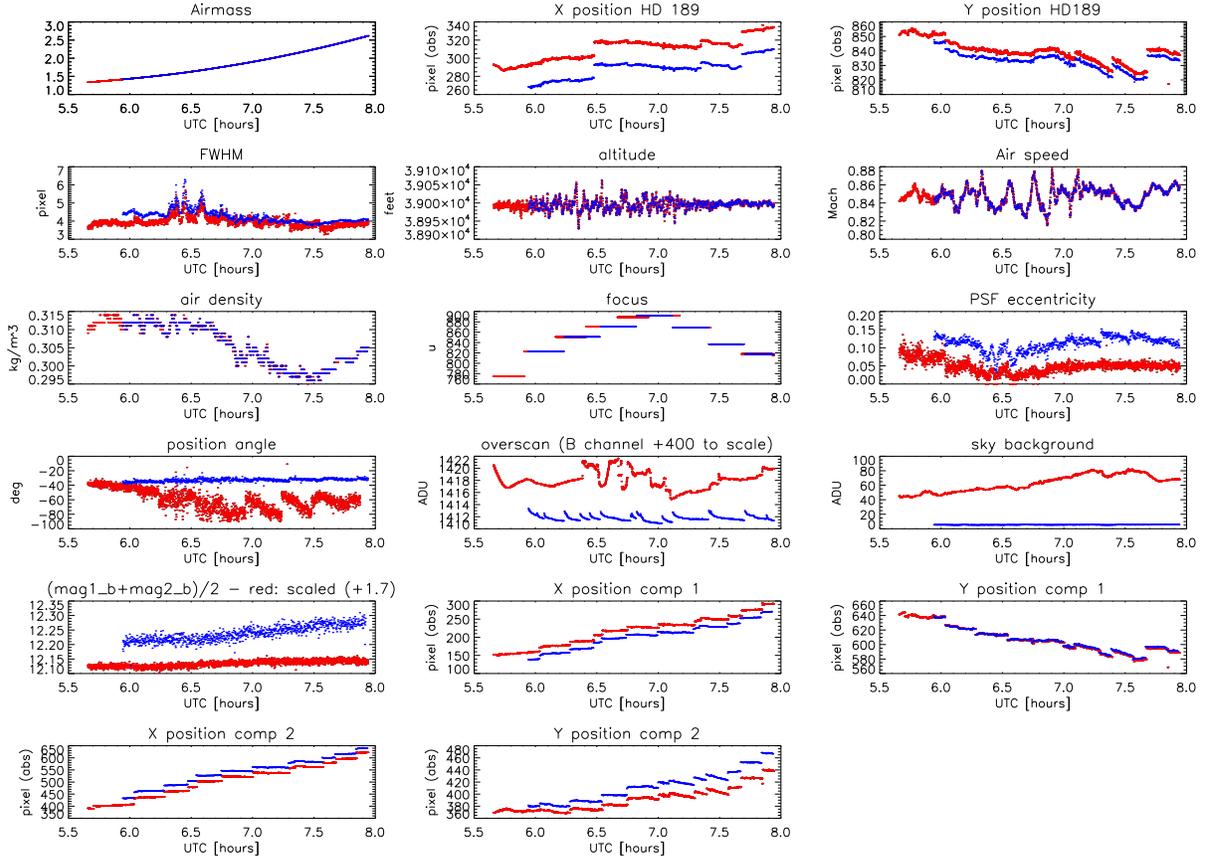}
       \caption{Time series of observational parameters for the B (blue) and $z^\prime$ (red) channel.  From top left to bottom right (for means and details about the decorrelation see Table \ref{tbl:corr1}) : airmass, X position of target star on detector, Y position, FWHM of PSF, altitude, mach number, density, focus, PSF eccentricity, overscan level, sky background, magnitude of comparison stars, X position of comparison star 1, Y position of comparison star 1, X position of comparison star 2 and Y position of comparison star 2. The position the comparison star has a different trend than the target since it rotates around the boresight, discontinuities in some parameters are cause by the unavoidable line of sight rewinds.  
}
     \label{fig:para_blue}
\end{figure*}


As an alternative we performed a principal component analysis (PCA) on this timeseries of observational parameters and used the principal components as vectors $p_i(t)$ in our modeling of the residual noise. For a PCA the vectors of the observational and environmental parameters (see Figure \ref{fig:para_blue}) that describe the temporal evolution of the observational environment (e.g., Mach number or flight altitude) or instrument behavior (PSF shape, detector position) are orthogonally transformed to convert this correlated set of variables into a set of values that are linearly uncorrelated -- the principal components. Advantages of using a PCA to disentangle the observational parameters are that it breaks degeneracies between parameters and reduces the number of fitting parameters. A disadvantage is loss of physical insight, since principal components are not always directly connectable to the original parameters.

The linear combination of observational parameters and PCA approaches were similarly effective in modeling the residual noise. The linear combination approach produced a slightly better rms of the final light curve so we used this method with the original vectors for the rest of our analysis.

\clearpage
 \begin{landscape}
\begin{deluxetable}{lcccccc}
\tablewidth{0pt}
\tablecaption{Correlation and fit coefficients of the observational parameters with the residuals\label{tab:corr1}}
\tablehead{\colhead{Parameter $p_i$}
& \colhead{Correl. (B)}  & \colhead{fit coeff. $c_i$ (B)}& \colhead{mean (B)}& \colhead{Correl. ($z^\prime$)}  & \colhead{fit coeff. $c_i$ ($z^\prime$)} & \colhead{mean ($z^\prime$)}}
\startdata

Airmass
&     -0.30

 & -0.003
 &1.91 
 &      0.01

& -0.007 

&1.847\\ 
 
X position HD189
&     -0.04

 &  0.017 
 
 &      288.66 px
& 0.02
 &  0.15 
 & 310.24 px\\ 
  
Y position HD189
&      0.30

 &   0.048 
 &833.130 px
 &     -0.12

  &     -0.155 
  & 839.67 px\\ 
   
FWHM
&      0.06

&  -0.009 
& 4.259 px
&      0.03

 &   -0.002  
 & 3.968 px\\

Altitude
&     -0.11

&   -0.122 
& 38995.1 ft 

&      0.03

 &    0.092 
 
 & 38994.7 ft\\   

Mach number
&      0.03

&   0.0155  

& 0.847

&     -0.11

&   -0.008   

& 0.848\\ 

 Density
&      0.17

 &  -0.036 
 
 & 0.305 $kg/m^3$
 
 &     -0.20

  &  -0.04 
  
  & 0.306 $kg/m^3$\\

 TA Focus
&     -0.12

& -0.009 

&852.61 u

&      0.46

&    0.006 

& 846.98 u\\ 

Image eccentricity
&     -0.09

 & -0.005 
 & 0.111
 &     -0.26

 &  -0.0002
 & 0.049\\ 
 
Position angle
&     -0.24

 &  0.0005 
 & -32.87 deg
 
 &     -0.06

 &  -0.00003
 & -56.683 deg\\  
 
Overscan
&      0.30

&  0.714  
& 1011.75 ADU

&      0.06

&   0.009
 & 1419.01 ADU\\ 

Sky Brightness
&     -0.36

&  -0.028 

& 5.45 ADU

&      0.15

&  -0.00006

& 62.68 ADU\\  

Comparison stars mag
&      0.03

 & -0.00007 
 
 &12.66 
 
 &     -0.00

 &  -0.081  
 
 & 10.43\\  
 
X pos, comparison 1
&     -0.23

 &  0.0025 
 
 & 202.637 pix
 
 &      0.13

&    -0.168 
& 216.338 pix\\ 
 
Y pos, comparison 1
&      0.10

 &  -0.018
 
 & 540.079 pix
 
 &     -0.09

 &      0.169   
 & 503.724 pix\\ 
 
X pos, comparison 2
&     -0.29

&  -0.0026  

&599.833 pix

&      0.12

  &  0.024  
  
  &607.72 pix\\ 

Y pos, comparison 1
&     -0.25

&   0.016

&411.21 pix

&      0.00

&    0.085
& 391.184 pix\\ 

\enddata
\label{tbl:corr1}
\end{deluxetable}

 \end{landscape}

\subsection{Analysis of a systematic effect during the transit baseline}\label{syseff}

\begin{figure*}[htp]
  \centering
      \includegraphics*[width=0.9\textwidth]{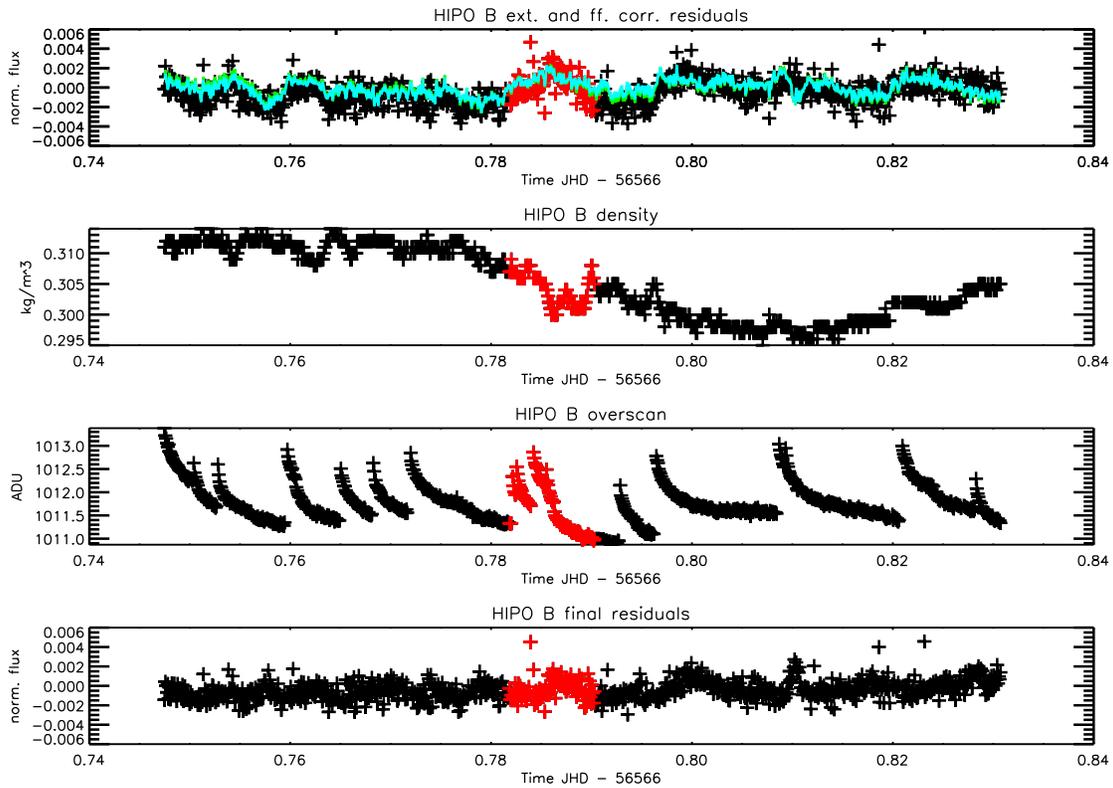}
       \caption{Top: residuals of the first order HIPO B lightcurve - the correlated feature (red) was excluded from first order noise models - a linear combination of observational parameters (green) or their principle components (light blue). Timeseries  of density (center top) and overscan (center bottom) illustrates the connection between the light curve feature and these observational parameters. Bottom: Final full residuals for HIPO B - in this case corrected by the green model extrapolated from the out-of-spot data. The potential spot signature (red) can almost completely be resolved by the noise model.
}
     \label{fig:ss_b}
\end{figure*}

\begin{figure*}[htp]
  \centering
      \includegraphics*[width=0.9\textwidth]{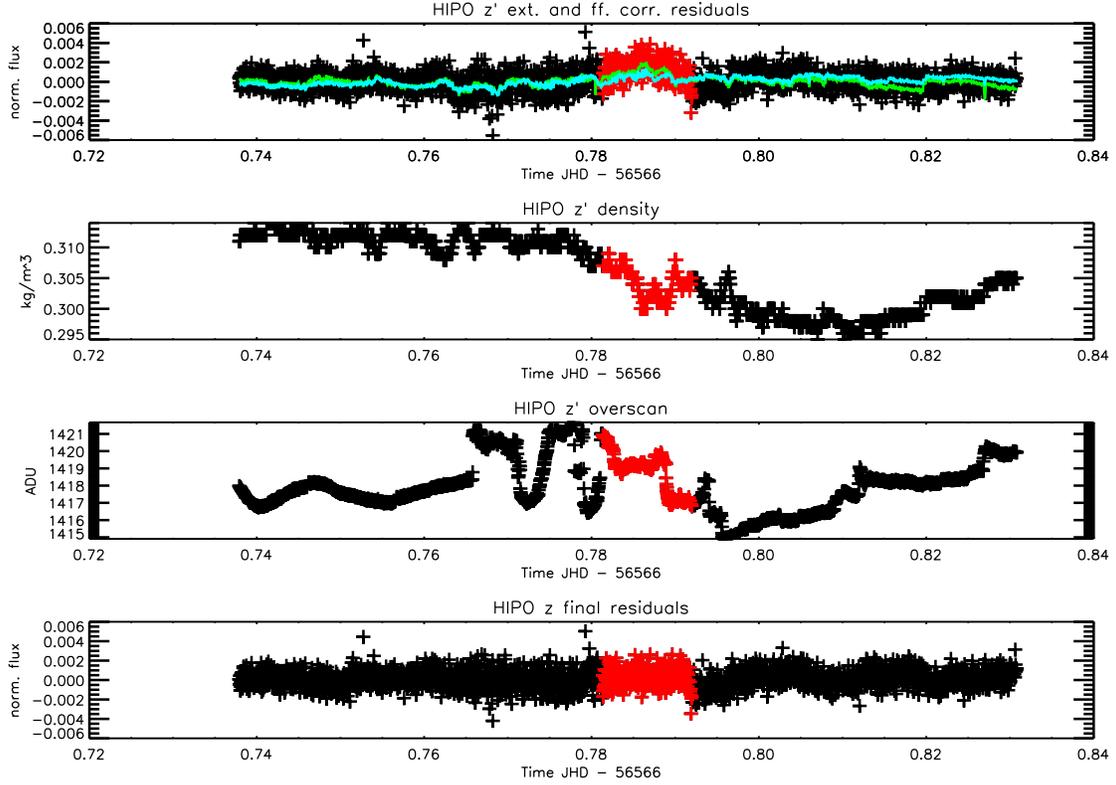}
       \caption{Same as Figure \ref{fig:ss_b} but for HIPO $z^\prime$.
}
     \label{fig:ss_z}
\end{figure*}

As mentioned before, we initially excluded a number of datapoints showing a systematic offset (that at first was identified as a potential starspot feature, red data points in Figure \ref{fig:1ord}) from our analysis, in order not to underestimate the resulting transit depth. In order to determine the source of this signature we looked for any potential dependence of the feature on our observational parameters. The density and overscan variations appeared to have a correlated behavior to the feature (see Figures \ref{fig:ss_b} and \ref{fig:ss_z}) and also occur during an extremal value of the focus parameter (see Figure \ref{fig:para_blue}). To test this, we interpolated the noise models from the fits to the masked-out data onto the data in the region of the potential starspot feature (light blue models in Figures \ref{fig:ss_z} and \ref{fig:ss_b}). As an alternative approach we applied an identical fit of a linear combination of the observational parameters to the unmasked residuals and then subtracted this noise model from the data. Using this method we were able to almost fully correct for the feature (green models in Figures \ref{fig:ss_b} and \ref{fig:ss_z}).
We can show that both models -- the one based on the full light curve as well as the model extrapolated from the fit to the masked data -- can correct for all of the spot in blue and most of it in red (see Figures \ref{fig:ss_z} and \ref{fig:ss_b}). A small signature in red remains, but the morphology still correlates with the overscan, density and focus parameters.

Considering the correlation with observational parameters and a wavelength dependence ($z\prime$ larger than B) that is opposite than expected from a starspot occultation. We therefore conclude that this light curve feature is most likely correlated with a non-linear dependence of the photometry with the observational parameters and is not caused by an astrophysical signal (as for example a starspot).

\subsection{MCMC Transit Fit }\label{section:MCMC}

To determine the uncertainty in the light curve parameters properly, we analyzed the corrected transit lightcurves with the IDL Transit Analysis Package (TAP) \citep{2012AdAst2012E..30G}. The TAP software utilizes Markov Chain Monte Carlo (MCMC) techniques to fit transit light curves using the \cite{2002ApJ...580L.171M} model.  TAP uses EXOFAST \citep{2013PASP..125...83E} to calculate the models. The package incorporates a wavelet-based likelihood function developed by \cite{2009ApJ...704...51C} which allows the MCMC algorithm to assess parameter uncertainties more robustly than classic $\chi^2$ methods by parameterizing uncorrelated `white"' and correlated `red' noise. However, we did not use the TAP red noise modeling in our reduction, but instead tried to model the correlated noise using our observational parameters (see discussion in Section \ref{noisemod}).

The planet-to-star radius ratio ($R_p/R_*$), the time of mid-transit ($T_0$) and the above-mentioned white noise and airmass parameters were the only free parameters in our fit.  We ran two iterations, one with the limb darkening parameters fixed to values from \cite{2011A&A...529A..75C} and the other with the limb darkening parameters as fitted parameters.

\cite{2013A&A...549A...9C} showed that fixing the limb darkening coefficients to theoretical values results in inaccuracies in the fitted planetary radius of more  than 1-10\%. They also demonstrated  how the presence of stellar spots can be responsible for this disagreement between the limb darkening tables and the fitted limb darkening coefficients, but showed that this does not affect the precision of the planet radius determination. Our results in comparison to previous observations confirm this finding. The recent analysis of \cite{2015arXiv150307020E} confirms that fixing the limb darkening coefficients can lead to significant biases up to a few percent and also conclude that, in this case, the best approach is to let them be free in the fitting procedure.

The scaled semi-major axis ($a/R_*$), the inclination (i)), the orbital period ($P$), the eccentricity ($e$) and the argument of periastron ($\omega$) were fixed to previously published values (e.g., Table \ref{tbl:blue_res}) because these parameters
are usually difficult to constrain using only a single transit for each target and are very accurately determined for HD~189733 b. For each transit light curve, we ran 5 MCMC chains of $10^6$ steps each.

\subsubsection{Analysis of the Full Lightcurve}

We showed that using a noise model, fitted as a linear combination of the observational vectors, enabled us to correct for most of the noise signature that looked like a potential starspot signature (see \ref{syseff}). Therefore we had to assume that it was most likely an instrumental effect and that we can also test our method on the whole data-set, including the potential starspot signature. For this analysis we used the same method as described above, but fitting our noise model and the transit lightcurve to all available data. Like the results shown in Figures \ref{fig:ss_b} and \ref{fig:ss_z} the full model was able to correct for the signature in the B channel but only partly in the $z^\prime$ channel. The fits to the full red lightcurve were clearly mislead by the above baseline values of this noisy signature (see e.g. Figure \ref{fig:ss_z}) leading to a much shallower estimate of the $z^\prime$ transit depth (see results in Table \ref{tbl:all_res}). Without more baseline data and corresponding housekeeping parameters it is hard to pin down the cause for this signature: since it strongly correlates with observational parameters (see Table \ref{tab:corr1} and Figures \ref{fig:ss_b} and \ref{fig:ss_z}) we suspect non-linear effects that were left uncorrected from our linear noise model.
In particular the TA focus value with a very high correlation value of 0.46 in $z^\prime$ (see Table \ref{tab:corr1}) and an extreme value during the time of the signature (see Figure \ref{fig:para_blue}, center) is known to have non-linear effects on the photometry.  Therefore we chose to prefer the results of the lightcurves with these data excluded for our futher analysis.

\section{Results}\label{chap:res}

\begin{figure*}[htp]
  \centering
      \includegraphics*[width=0.9\textwidth]{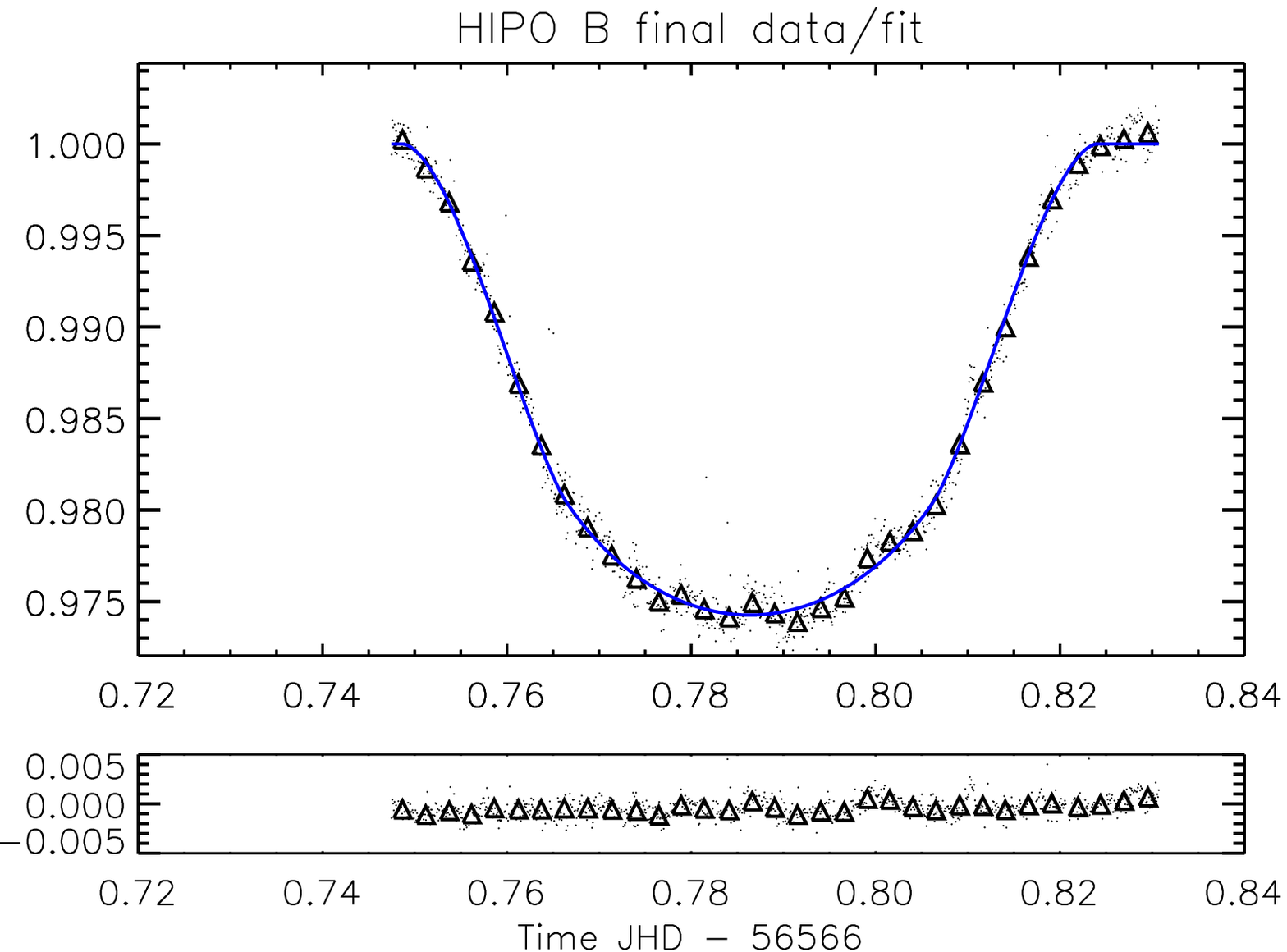}
       \caption{Final lightcurve in HIPO B, fit (blue, with free limb darkening). Triangles are averaged over 30 datapoints. For derived precision (also for the binned data) see Table \ref{tbl:noisecomp}.
}
     \label{fig:fin_b}
\end{figure*}

\begin{figure*}[htp]
  \centering
      \includegraphics*[width=0.9\textwidth]{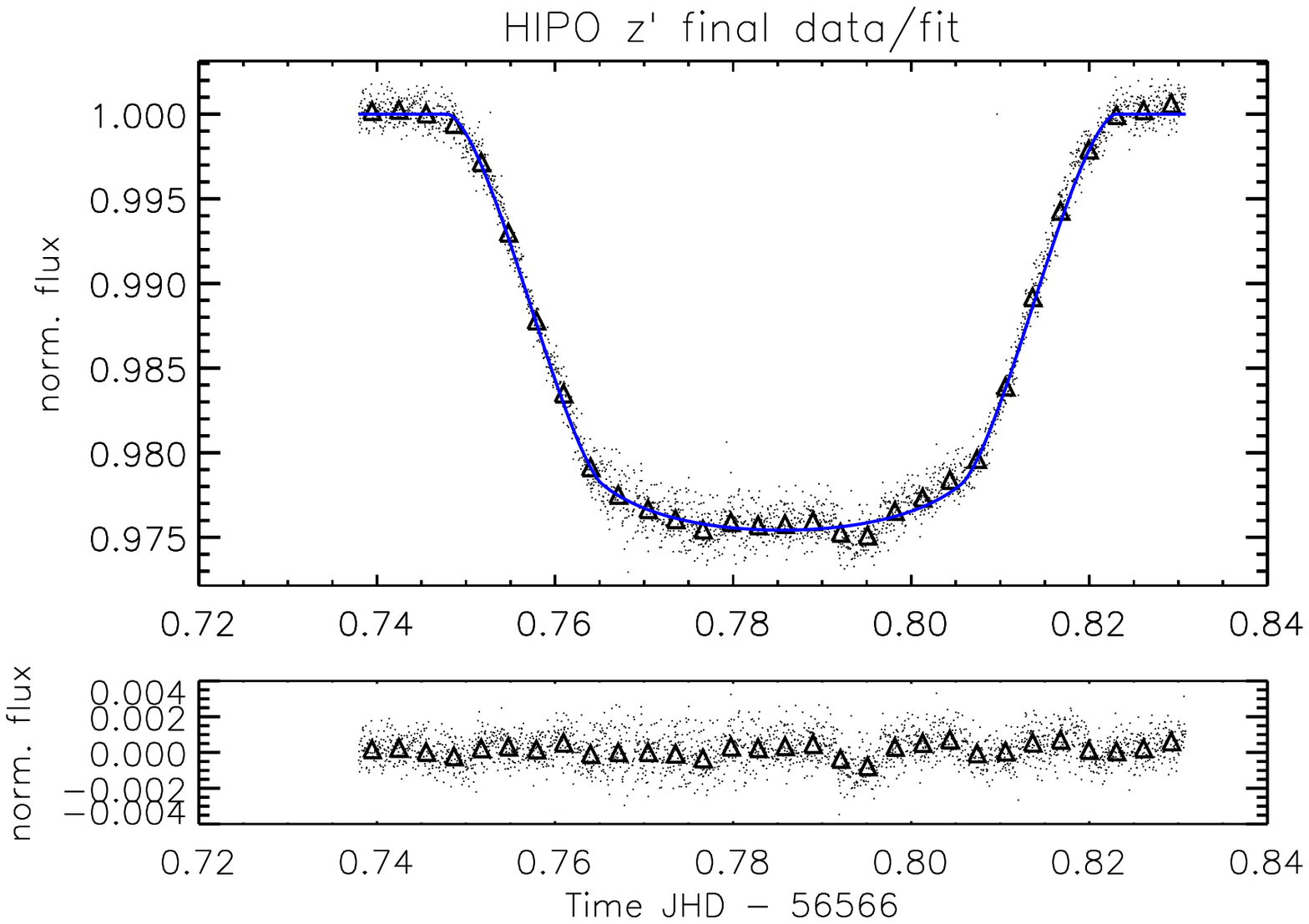}
       \caption{Final lightcurve in HIPO $z^\prime$, fit (blue, with free limb darkening). Triangles are averaged over 87 datapoints. For derived precision (also for the binned data) see Table \ref{tbl:noisecomp}.
}
     \label{fig:fin_z}
\end{figure*}

\begin{figure*}[htp]
  \centering
      \includegraphics*[width=0.9\textwidth]{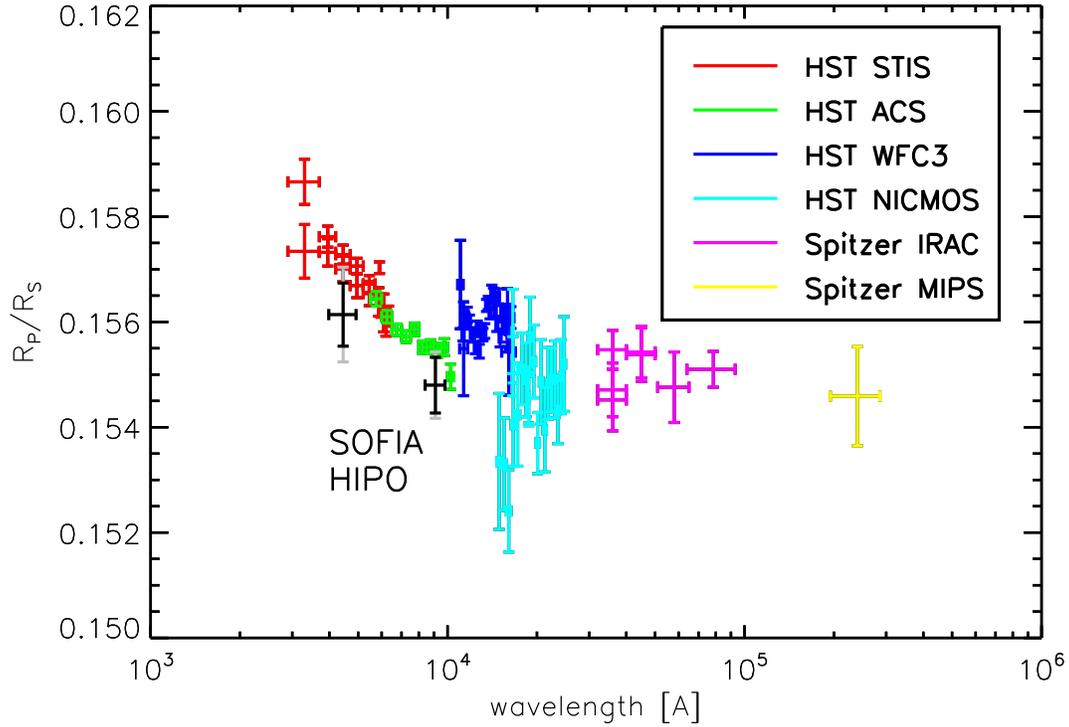}
       \caption{Fit results of our HIPO-SOFIA observation (black: after noise correction, grey: maximum uncertainties of noise model)in comparison to previous observations from other space-based platforms such as \textit{HST} and \textit{Spitzer} \citep[see legend, from ][]{2013MNRAS.432.2917P,2014ApJ...791...55M}. The  error bars are given by the TAP MCMC routine. Our B and $z^\prime$ data points reproduce the slope in the optical, most likely caused by Rayleigh scattering, though with an overall smaller transit depth, with might be caused by a difference in unocculted starspots between the observed epochs \citep{2014ApJ...791...55M}. Given that our observation (1) had very little out of transit baseline, (2) taken with an only $30\%$ reflective tertiary and that (3) there is still room for improvement for the correction of systematics once they are better understood with further observations, we can report that SOFIA should be able to deliver space based data quality at least in the optical with HIPO. The goal for the future is to reach similar quality in the infrared with FLITECAM, to cover a wavelength region that will not be available after the end of \textit{Spitzer} and the decommissioning of NICMOS until the start of JWST.} 
     \label{fig:comp_res}
\end{figure*}

Our fit results for the corrected HIPO B lightcurves (see Figure \ref{fig:fin_b}) are shown in Table \ref{tab:b2}, and our fit results for the corrected HIPO $z^\prime$ lightcurves (see Figure \ref{fig:fin_z}) are shown in Table \ref{tab:z1}.  We find transit depths in B and $z^\prime$ that are consistent with previous observations (see Figure \ref{fig:comp_res}) with \textit{HST}'s ACS and STIS instruments and confirm the slope in the optical most likely due to some form of Rayleigh scattering.   


We believe the accuracy of future transit observations with SOFIA will be significantly improved compared with our current results (more in Section \ref{diss:noise}): Our observation suffered from a lack of out-of-transit baseline, decreasing the overall precision of the transit measurement, and there is also room for improvement in the understanding of the systematics caused by the airborne environment. Additionally, the observations were further degraded due to the low reflectivity ($30\%$) of the current SOFIA tertiary mirror.  Future observations with HIPO will seek to improve upon the first two areas, and we hope to reach a similar precision in the infrared with FLITECAM.

    
\begin{deluxetable}{lcccc}
\tablewidth{0pt}
\tablecaption{MCMC fit results for the HIPO B lightcurve, `starspot' excluded \label{tab:b2}}
\tablehead{
\colhead{Parameter}
& \colhead{raw, fixed LD} & \colhead{raw, free LD}
& \colhead{corr, fixed LD} & \colhead{corr, free LD}
}
\startdata
         Period (d)& 2.2185741\tablenotemark{a}& 2.2185741\tablenotemark{a}& 2.2185741\tablenotemark{a}& 2.2185741\tablenotemark{a}\\
    Inclination ($^{\circ}$)& 85.591\tablenotemark{a}& 85.591\tablenotemark{a}& 85.591\tablenotemark{a}& 85.591\tablenotemark{a}\\
           a/R* & 8.8251\tablenotemark{a}& 8.8251\tablenotemark{a}& 8.8251\tablenotemark{a}& 8.8251\tablenotemark{a}\\
            Eccentricity & 0.\tablenotemark{a}& 0.\tablenotemark{a}& 0.\tablenotemark{a}& 0.\tablenotemark{a}\\
          Omega ($^{\circ}$)& 0.\tablenotemark{a}& 0.\tablenotemark{a}& 0.\tablenotemark{a}& 0.\tablenotemark{a}\\ \hline
          $R_p/R_*$ & 0.15585 $^{+0.00053}_{-0.00054}$& 0.15686 $^{+0.00061}_{-0.00064}$& 0.15527 $^{+0.00040}_{-0.00041}$& 0.15614 $^{+0.00052}_{-0.00060}$\\
    Mid Transit & 0.78631 $^{+0.00011}_{-0.00011}$& 0.78629 $^{+0.00011}_{-0.00011}$& 0.786511 $^{+0.000082}_{-0.000081}$& 0.786485 $^{+0.000084}_{-0.000084}$\\
      Linear LD & 0.8541\tablenotemark{b}& 0.978 $^{+0.017}_{-0.036}$& 0.8541\tablenotemark{b}& 0.967 $^{+0.024}_{-0.049}$\\
        Quad LD & -0.00921\tablenotemark{b}& -0.155 $^{+0.047}_{-0.030}$& -0.00921\tablenotemark{b}& -0.138 $^{+0.064}_{-0.034}$\\
  
  Airmass Y-int & 0.99935 $^{+0.00021}_{-0.00021}$& 0.99940 $^{+0.00024}_{-0.00023}$& 0.99902 $^{+0.00016}_{-0.00016}$& 0.99905 $^{+0.00019}_{-0.00018}$\\
  Airmass Slope & 0.0103 $^{+0.0036}_{-0.0036}$& 0.0097 $^{+0.0037}_{-0.0037}$& 0.0161 $^{+0.0028}_{-0.0027}$& 0.0153 $^{+0.0029}_{-0.0029}$\\
      Sigma Red & 0.\tablenotemark{c}& 0.\tablenotemark{c}& 0.\tablenotemark{c}& 0.\tablenotemark{c}\\
    Sigma White & 0.001145 $^{+0.000037}_{-0.000035}$& 0.001137 $^{+0.000037}_{-0.000035}$& 0.000879 $^{+0.000028}_{-0.000027}$& 0.000877 $^{+0.000029}_{-0.000027}$\\
\enddata
\tablenotetext{a}{Value Fixed from \cite{2008ApJ...677.1324T}}
\tablenotetext{b}{Value Fixed from \cite{2011A&A...529A..75C}}
\tablenotetext{c}{Fixed in first fit in order to model red noise independently}
\label{tbl:blue_res}
\end{deluxetable}

\begin{deluxetable}{lcccc}
\tablewidth{0pt}
\tablecaption{MCMC fit results for the HIPO $z^\prime$ lightcurve, `starspot' excluded \label{tab:z1}}
\tablehead{
\colhead{Parameter}
& \colhead{raw, fixed LD} & \colhead{raw, free LD}
& \colhead{corr, fixed LD} & \colhead{corr, free LD}
}
\startdata
         Period (d)& 2.2185741\tablenotemark{a}& 2.2185741\tablenotemark{a}& 2.2185741\tablenotemark{a}& 2.2185741\tablenotemark{a}\\
    Inclination ($^{\circ}$)& 85.591\tablenotemark{a}& 85.591\tablenotemark{a}& 85.591\tablenotemark{a}& 85.591\tablenotemark{a}\\
           a/R* & 8.8251\tablenotemark{a} & 8.8251\tablenotemark{a}& 8.8251\tablenotemark{a} & 8.8251\tablenotemark{a}\\
             Eccentricity & 0.\tablenotemark{a}& 0.\tablenotemark{a}& 0.\tablenotemark{a}& 0.\tablenotemark{a}\\
          Omega ($^{\circ}$) & 0.\tablenotemark{a}& 0.\tablenotemark{a}& 0.\tablenotemark{a}& 0.\tablenotemark{a}\\ \hline
          
          $R_p/R_*$\tablenotemark{c} & 0.15431 $^{+0.00018}_{-0.00018}$& 0.15315 $^{+0.00054}_{-0.00046}$& 0.15418 $^{+0.00017}_{-0.00018}$& 0.15355 $^{+0.00053}_{-0.00049}$\\
    Mid Transit & 0.785441 $^{+0.000033}_{-0.000033}$& 0.785460 $^{+0.000033}_{-0.000033}$& 0.785504 $^{+0.000031}_{-0.000033}$ & 0.785520 $^{+0.000031}_{-0.000031}$\\
      Linear LD & 0.349\tablenotemark{b}& 0.133 $^{+0.089}_{-0.082}$& 0.349\tablenotemark{b} & 0.219 $^{+0.087}_{-0.083}$\\
        Quad LD & 0.221\tablenotemark{b}& 0.412 $^{+0.094}_{-0.10}$& 0.221\tablenotemark{b}& 0.285 $^{+0.096}_{-0.10}$\\
 
  Airmass Y-int & 0.999910 $^{+0.000057}_{-0.000057}$ & 0.999950 $^{+0.000059}_{-0.000060}$& 0.999834 $^{+0.000050}_{-0.000053}$& 0.999929 $^{+0.000056}_{-0.000056}$\\
  Airmass Slope & 0.0017 $^{+0.0010}_{-0.00099}$ & 0.00232 $^{+0.00099}_{-0.00099}$& 0.00311 $^{+0.00091}_{-0.00097}$& 0.00382 $^{+0.00094}_{-0.00093}$\\
      Sigma Red & 0.\tablenotemark{d}& 0.\tablenotemark{d}& 0.\tablenotemark{d}& 0.\tablenotemark{d}\\
    Sigma White & 0.000728 $^{+0.000011}_{-0.000011}$& 0.000721 $^{+0.000011}_{-0.000011}$& 0.000694 $^{+0.000011}_{-0.000010}$& 0.000681 $^{+0.000011}_{-0.000010}$\\
\enddata
\tablenotetext{a}{Value Fixed from \cite{2008ApJ...677.1324T}}
\tablenotetext{b}{Value Fixed from \cite{2011A&A...529A..75C}}
\tablenotetext{c}{0.0012 too low at $z^\prime$ due to companion contamination}
\tablenotetext{d}{Fixed in first fit in order to model red noise independently}
\label{tbl:red_res}
\end{deluxetable}

\begin{deluxetable}{lcccc}
\tablewidth{0pt}
\tablecaption{MCMC fit results for the full HIPO lightcurve, systematic outlier  included \label{tab:incl}}
\tablehead{
\colhead{Parameter\tablenotemark{a} }
& \colhead{HIPO B} & \colhead{HIPO z}
}
\startdata

         $R_p/R_*$ & 0.15607 $^{+0.00054}_{-0.00059}$& 0.15275 $^{+0.00025}_{-0.00022}$\tablenotemark{b}\\

    Mid Transit & 0.786505 $^{+0.0001}_{-0.000033}$& 0.785545 $^{+0.00003}_{-0.00003}$&\\

      Linear LD & 0.969  $^{+0.022}_{-0.047}$ & 0.033 $^{+0.043}_{-0.024}$\\

        Quad LD & -.133  $^{+0.063}_{-0.033}$ & 0.514 $^{+0.029}_{-0.05}$\\
 
  Airmass Y-int & 0.99903 $^{+0.00021 }_{-0.00021 }$ & 0.999981 $^{+0.00006 }_{-0.00006 }$\\

  Airmass Slope & 0.014  $^{+0.003 }_{-0.003  }$ & 0.004   $^{+0.001  }_{-0.001  }$\\

  Sigma White & 0.001    $^{+0.00003 }_{-0.00003 }$& 0.000768 $^{+0.00001 }_{-0.00001 }             $\\

\enddata
\tablenotetext{a}{All other values fixes from \cite{2008ApJ...677.1324T}}

\tablenotetext{b}{0.0012 too low at $z^\prime$ due to companion contamination}

\label{tbl:all_res}
\end{deluxetable}

\subsection{Comparison to Expected Photon Noise}\label{fig:comp_noise}

We performed a calculation of the expected photon noise levels based on our observed star and sky signal levels. We find a signal in electrons per exposure of $5.81 \times 10^6$ in the B channel,and  $1.44\times 10^7$ in  $z^\prime$. The total sky contribution in this aperture in electrons is 
$1.46\times 10^5$ for B, and $2.83\times 10^6$ in $z^\prime$. The total noise was computed by taking the square root sum of the shot noise values of the stellar signal (2410 $e^-$ B, 3800$e^-$ $z^\prime$), the  shot noise contribution of the sky background (380 $e^-$ B, 1680 $e^-$ $z^\prime$), and a read noise ($6 e^-$ per pixel) of 975 $e^-$ on both sides.  

This gives a total noise of 2600 $e^-$ in B and 4300 $e^-$ in $z^\prime$, resulting in a final
photon noise limited signal-to-noise ratio (SNR) of 2200 per 7 second integration in B, and 3400 per 3 second integration in $z^\prime$.
This is very close to the theoretically derived values of a SNR of 2010 in B and 3990 in $z^\prime$ from the HIPO exposure time calculator spreadsheet (provided by the HIPO instrument PI \footnote{http://www.sofia.usra.edu/Science/ObserversHandbook}). Most of the difference is due to the sky contribution, which was unassessed prior to the observations. With our observation we were able to obtain real data on the sky background at SOFIA's flight altitude, which will be used to improve future exposure time estimates.

Adding an allowance for scintillation noise (a poorly understood contribution that may be important for bright stars like HD 189733 - here we used the recipe from the aforementioned HIPO exposure time calculator spreadsheet that is based on the findings in \citep{2012SPIE.8446E..18D}), the derived SNR values are adjusted downward to about 1900 in B and 2800 in $z^\prime$. We used these numbers for comparison and also to model photon noise limited lightcurves to test the MCMC fitting routine. Indeed, we find that the limited amount of out of transit data increases the error by a factor of $1.5$. The final photometric precision (600 ppm in B and 530 ppm in $z^\prime$ for the fitted parameter $(R_p/R_*)$, see Figure \ref{fig:comp_res}, corresponding to 187 ppm in B and 165 ppm in $z^\prime$ for the actual transit depth $(R_p/R_*)^2$ respectively) is close to two times the expected photon noise (see error bars in Figure \ref{fig:comp_res}). While our initial noise model(\ref{noisemod}) was able to reduce the raw standard deviation $\sigma_{raw}$ by about 50\% in B and 30\% in $z^\prime$ (see Table \ref{tbl:noisecomp}), we are still not at the photon noise lever yet. As already discussed briefly in \ref{syseff} we suspect non-linearities, in particular in highly correlated observational parameters  such as overscan, density or TA focus (see Table \ref{tab:corr1} and Figures \ref{fig:ss_b} and \ref{fig:fin_z}) to be responsible for this residual correlated noise.

%

\begin{deluxetable}{lcccccc}
\tablewidth{0pt}
\tablecaption{Error analysis. Resulting precision for both channels: fit parameter $\Delta (R_p/R_*)$ - theoretical value and from EXOFAST MCMC; standard deviations for the binned data (from Figures \ref{fig:fin_b} and \ref{fig:fin_z}), the final data after the slope correction, and the raw lightcurves; final errors for the transit depth $\Delta (R_p/R_*)^2$ }
\tablehead{\colhead{}& \colhead{$\Delta (R_p/R_*)$} & \colhead{}&\colhead{stand. dev.}& \colhead{}  & \colhead{}& \colhead\textbf{$\Delta (R_p/R_*)^2$ }
\\
\colhead{Channel}& \colhead{theoretical} & \colhead{from MCMC}&\colhead{$\sigma_{binned}$}& \colhead{$\sigma_{final}$}  & \colhead{$\sigma_{raw}$}& \colhead\textbf{from MCMC}
}
\startdata

         B & 0.00051 & 0.00060 & 0.00046 & 0.00102 & 0.00157& \textbf{187 ppm}\\
        $z^\prime$ & 0.00036 & 0.00053& 0.00033 & 0.00094 & 0.00118 & \textbf{165 ppm} \\
        
\enddata
\label{tbl:noisecomp}
\end{deluxetable}

\section{Summary and Conclusions}\label{chap:sum}

The results of our observation can be summarized as follows:
We demonstrate absolute photometry in the airborne environment with the HIPO instrument and show that SOFIA is a quasi-space based platform  for photometry in the optical. We have a reasonable understanding of the systematic noise of photometric observation in SOFIA's airborne environment, but there is still room for improvement.  The biggest improvement would be a larger amount of baseline data to use for deriving the systematic error corrections.

\subsection{HD 189733 b}

A study of \cite{2013MNRAS.432.2917P} provided the transmission spectrum across the entire visible and infrared range in a consistent analysis, including the general system parameters and stellar limb darkening. 
Their resulting spectrum does not confirm the predicted  cloud-free atmospheres of highly-irradiated hot Jupiters. Instead they show a Rayleigh scattering slope over the whole visible and near-infrared range, with only narrow $Na$ and $K$ lines detected.  \citep{2014ApJ...791...55M}, however, argued that the slope between the visible and near-IR shown in Figure 9 can partly be caused by unocculted starspots and not Rayleigh scattering alone.

We were able to confirm this previous reported slope \citep{2013MNRAS.432.2917P}, that can be explained  by a haze of condensate grains, in the upper atmosphere of HD~189733 b  (see Figure \ref{fig:comp_res}). The slight overall offset between the SOFIA and HST observations, however, could indeed be explained by a difference in starspot coverage between the epochs and confirm the \citep{2014ApJ...791...55M} interpretation that the slope is caused by unocculted starspot combined with Rayleigh scattering from molecular hydrogen only instead of dust. Furthermore a difference in spot coverage can also explain the difference between the fitted free limb darkening values from the fixed theoretical values and the divergent resulting transit depths in the corresponding fits \cite{2013A&A...549A...9C,2015arXiv150307020E}. 

\subsection{Comparison to other platforms}\label{diss:noise}
 
 Our transit observation of HD 189733 b with SOFIA-FLIPO cover some of the bands that were observed with other platforms (\textit{HST} and \textit{Spitzer}). This enables us to compare SOFIA with the best existing observatories, paving the way for future observations of a wide range of exoplanet atmospheres with FLIPO on SOFIA.

The comparison of the sensitivity of our observation (187 ppm in B and 165 ppm in $z^\prime$ for the  transit depth $(R_p/R_*)^2$) with similar observations with \textit{HST} and \textit{Spitzer} (see Figure \ref{fig:comp_res}) show that SOFIA is a competitive platform in this field, in particular if viewed next to to the very first exoplanet observations with these platforms \cite{2002ApJ...568..377C, 2005Natur.434..740D,2005ApJ...626..523C}. Comparing these early proof of concept observations with this very first SOFIA exoplanet observation illustrates this. In detail the room for improvement is threefold:
(1) as described earlier the observatin was limited to less than a quarter of an hour out of transit baseline, about a factor 6-8 less data then actually expected (2) only about $30\%$ of the light was passed to HIPO by the reflective tertiary, which provides room for about 3 times more signal (3) lessons learned from this and upcoming exoplanet observations (PIs: Angerhausen/02-0053, 03-0052, Dreyer/02-0084, Huber/03-0042 and Swain/03-0037) will further improvement our understanding for the correction of systematic noise.

\subsection{Future prospects with SOFIA}

 The main goal for the future has to be to reach similar quality in the infrared with FLITECAM. This will cover a wavelength region that will not be available after the end of \textit{Spitzer} and the decommissioning of NICMOS until the start of JWST. In particular the planned observation to demonstrate FLITECAM's spectroscopic transit mode (PI: Swain/03-0037) will hopefully pave the way for this.

A fully reflective tertiary and speciality filters for HIPO (e.g. special filters to avoid O$_{3}$ and H$_{2}$O, or to analyse narrow band $K$ or $Na$) will further improve SOFIA's capabilities to characterize exoplanets. In contrast to space-telescopes it is possible and highly desired to update SOFIA with state-of-the-art instrumentation. With regard to upcoming calls for next generation instruments, we support the idea of equipping SOFIA with a dedicated 2$^{nd}$ generation exoplanet instrument \citep[e.g., NIMBUS][]{2012SPIE.8446E..7BM}. Another important point is to ensure better baseline coverage for future observation, e.g. by allowing special deployments for these kind of time critical observations.

In the context of other platforms SOFIA has obvious synergies with ground-based telescopes: SOFIA is `filling the gaps' between bands that are unobservable from the ground due to mostly H$_2$O absorption. These are crucial bandpasses to characterize   important molecules (H$_2$O, CH$_4$, CO$_2$, PAHs etc) in exoplanetary atmospheres. In particular after the rejection of characterization proposals for dedicated space observatories (like EChO or FINESSE), SOFIA is -- until the start of JWST -- the only quasi space based platform $>$ 1.7 micron (\textit{HST} WFC3 limit). 

%

\acknowledgments 

We thank the whole SOFIA team, in particular the flight planners. DA's research was supported by 
an appointment to the NASA Postdoctoral Program at the Goddard Space Flight Center, administered by Oak Ridge Associated Universities through a contract with NASA.  HIPO work at Lowell is supported by USRA subcontract 8500-98-003. FLITECAM work at UCLA is supported by USRA subcontract 08500-05, PI Ian McLean.
We thank the anonymous referees for their detailed review, that significantly helped to improve the quality of this publication.


\bibliographystyle{spiejour}   


\vspace{2ex}\noindent{\bf Daniel Angerhausen} is a NASA Postdoctoral Program Fellow at NASA Goddard Space Flight Center. He received his diploma degrees in physics from the University of Cologne in 2006, and his PhD degree in physics from the German SOFIA Institute, University Stuttgart in 2010. 

\vspace{1ex}
\noindent Biographies and photographs of the other authors are not available.

\listoffigures
\listoftables

\end{spacing}
\end{document}